\documentclass[letterpaper, 10 pt, journal, twoside]{ieeetran}

\IEEEoverridecommandlockouts                              

\usepackage{tcolorbox}
\usepackage{listings}
\usepackage{xcolor}
\usepackage{cite}
\usepackage{enumitem}
\definecolor{lightgraybg}{RGB}{245,245,245}
\definecolor{bordergray}{RGB}{180,180,180}
\definecolor{stringgreen}{RGB}{120,180,100}
\definecolor{lineblue}{RGB}{220,235,248}
\usepackage[normalem]{ulem}
\usepackage{booktabs}
\usepackage{multirow}
\usepackage{amsmath}
\usepackage{fancyhdr}
\usepackage{amssymb}
\usepackage{float} 
\usepackage{moresize}
\usepackage{hyperref}
\usepackage{threeparttable}
\lstdefinestyle{promptstyle}{
    breaklines=true,
    breakindent=0pt,
    breakautoindent=false,
    backgroundcolor=\color{lightgraybg},
    basicstyle=\ttfamily\fontsize{6}{7}\selectfont,
    frame=single,
    rulecolor=\color{bordergray},
    framerule=0.8pt,
    xleftmargin=4pt,
    xrightmargin=4pt,
    aboveskip=3pt,
    belowskip=3pt,
    columns=fullflexible,
    keepspaces=true,
    showstringspaces=false,
    breaklines=true,
    breakatwhitespace=false,
    stringstyle=\color{stringgreen},
    keywordstyle=\color{black},
    commentstyle=\color{black},
    escapeinside={(*@}{@*)}
}

\author{Pourya Aliasghari and Goldie Nejat, \textit{Member, IEEE}
\thanks{
This work was supported by the Canadian Institute for Advanced Research~(CIFAR) Innovation, Equity \& the Future of Prosperity~(IEP) program, the Canada Research Chairs~(CRC) program, the Canadian Foundation for Innovation~(CFI), the Ontario Research Fund, and the Natural Sciences and Engineering Research Council of Canada (NSERC). \textit{(Corresponding author: Pourya Aliasghari)}} 
\thanks{The authors are with the Autonomous Systems and Biomechatronics Laboratory (ASBLab), Department of Mechanical and Engineering, University of Toronto, Toronto, ON M5S 3G8, Canada (e-mail:
        {\tt\footnotesize pourya.aliasghari@utoronto.ca, \tt\footnotesize goldie.nejat@utoronto.ca})}
}

\title{
PARAssist: A Framework for Personalized and Adaptive Robotic Assistance from Ambiguous User Requests}

\begin{document}
\maketitle

\begin{abstract}

Service robots may encounter ambiguous user requests that require context-aware inference. Users may also have unique preferences with certain tasks when requesting robotic assistance. We introduce PARAssist (Personalized and Adaptive Robotic Assistance), a unique architecture for disambiguating requests in a personalized manner for service robots. PARAssist utilizes vision-language models to determine the physical and cognitive demands of a user's tasks, and passively learns user preferences for assistance by contrasting the demands of tasks the user performs independently with those they request from the robot. When an ambiguous request is received, task candidates are generated from the history of the user's actions, activities, locations, conversations, and requests, as well as the current user and environment state. Task candidates are then evaluated against the learned user preference model to suggest suitable assistance options. Experiments conducted with PARAssist show that personalization can align disambiguation with the task demands of a user's prior assistance requests. An ablation study confirms the contributions of PARAssist's main components in personalizing disambiguation.

\end{abstract}
\vspace{-0pt}
\vspace{-0pt}
\section{Introduction}

\IEEEPARstart{S}{ervice} robots can assist users with a variety of tasks from household chores~\cite{Wu2023} to hospital medication delivery~\cite{Kostavelis2022}. 
However, verbal task requests can be ambiguous and may not provide key details such as where exactly to place items~\cite{Ren2023} or which items to retrieve~\cite{Pramanick2022}. 
In human-human interactions, 
a requester's intent can be inferred from common social context~\cite{Gibbs1985}, prior interactions, and/or knowledge of the person and the environment~\cite{Abbeduto1989}. Furthermore, a user’s specific capabilities and preferences may be considered, namely, if they have physical or cognitive impairments. 

For robots to provide appropriate assistance when task requests are ambiguous, methods such as fuzzy logic~\cite{Muthugala2016}, semantic parsers~\cite{Thomason2015}, image captioning networks~\cite{Dogan2022,Pramanick2022}, Large Language Models~(LLMs)~\cite{Ren2023,Liang2024, Park2024,Abugurain2024}, and Vision-Language Models~(VLMs)~\cite{Chisari2025} have been used. These methods resolve ambiguity through either: 1)~user intent inference~\cite{Muthugala2016,Wan2025}, 2)~clarification questions~\cite{Ren2023,Thomason2015,Dogan2022,Pramanick2022,Chisari2025,Park2024,Abugurain2024}, or 3)~candidate task suggestions based on the current scene~\cite{Liang2024}. 
Service robots should also provide personalized assistance to individual users based on their needs and wants~\cite{Smarr2014}. 
The aforementioned approaches for disambiguation use general-purpose models, which are not able to resolve ambiguity based on unique and evolving user preferences~\cite{Ren2023,Liang2024,Park2024,Abugurain2024,Chisari2025}. This can limit the adaptability of service robots to diverse users. Furthermore, methods that exist for personalizing robotic assistance mainly focus on learning user preferences for \textit{how} tasks should be performed and rely on explicit user-provided examples~\cite{Sadigh2017,Zhang2019,Wu2023,Wang2025,Han2025}, rather than inferring preferences from naturalistic user behavior. Demonstrating preferences to robots may require users to perform the tasks themselves, which can be challenging, particularly for older adults with limited mobility~\cite{Smarr2014,Czaja2006}.

\begin{figure}[tp]
    \centering
    \includegraphics[width=0.96\linewidth]{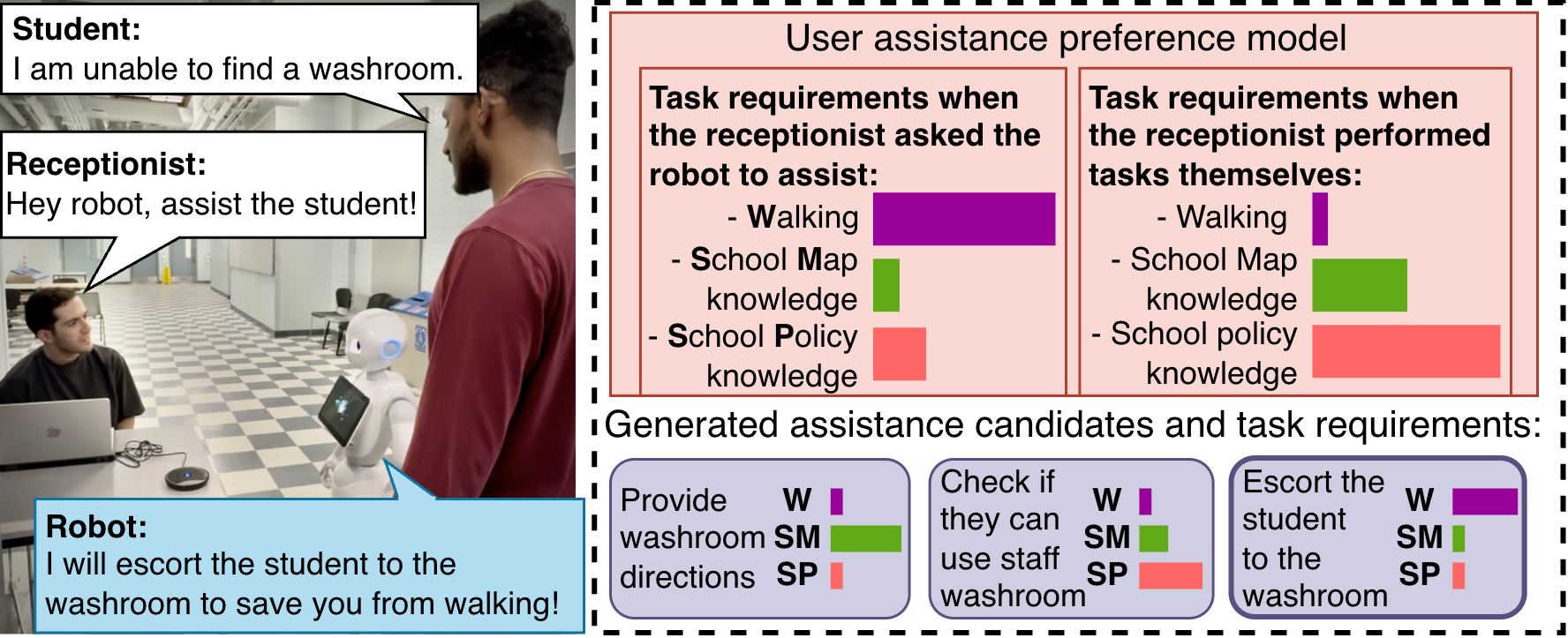}\vspace{-0pt}
    \caption{Overview of PARAssist functionality. The robot learns the preferences of a user (e.g., a receptionist at a university) for receiving robotic assistance based on VLM-inferred task requirements. PARAssist generates candidate tasks and evaluates them using a user assistance preference model. In this example, in response to the user’s ambiguous request of \textit{“Hey robot, assist the student,”} the robotic assistance of `escorting the student to the washroom' was suggested as it had the highest likelihood based on the observed patterns in requesting robotic assistance when walking is required.}
    \label{fig:banner}
    \vspace{-0pt}
\end{figure}

To address the challenges of enabling robots to disambiguate user requests in a personalized manner by passively learning user preferences for assistance, we present the Personalized and Adaptive Robotic Assistance~(PARAssist) architecture.
PARAssist utilizes VLMs to reason over the context of a task scenario and the history of user states to infer candidate tasks to assist with. 
PARAssist uniquely integrates a user assistance preference model into a disambiguation pipeline to suggest the highest likelihood task candidate for the specific user, Fig.~\ref{fig:banner}. 
Its novelty lies in combining passive, multimodal observations of user actions, activities, locations, and conversations with a constantly updated user preference model to personalize disambiguation.

Our main contributions are the development of: 1) a VLM-based assistance candidate generation module to disambiguate user requests and produce a set of context-relevant candidate tasks based on both the current task scenario and past user states, and 2) a User Assistance Preference Model~(UAPM), which is learned passively without explicit user demonstrations or feedback. Namely, the VLM-inferred physical and cognitive demands of user task performance and user requests from the robot are used for training the UAPM. PARAssist determines suitable assistance by combining a user preference score with a relevance score for the scenario context and the history of user states.

\section{Related Work}

We discuss previous work on how service robots: 1) handle ambiguous user requests~\cite{Thomason2015,Chisari2025,Dogan2022,Pramanick2022,Wan2025,Park2024,Ren2023,Muthugala2016,Abugurain2024,Liang2024}, and 2) personalize their assistance~\cite{Sadigh2017,Zhang2019,Wu2023,Wang2025,Han2025,Candon2026,Patel2025}.

\vspace{-0pt}
\subsection{Service Robots Handling Ambiguous Requests}

When users provide ambiguous or underspecified natural language requests, service robots need to resolve ambiguities related to: 1) object selection~\cite{Thomason2015,Chisari2025,Dogan2022,Pramanick2022,Wan2025,Park2024}; 2) locations, directions, and spatial relations~\cite{Thomason2015,Ren2023,Muthugala2016,Abugurain2024}; and/or 3) the intended type of assistance needed~\cite{Liang2024,Park2024}. 
Once ambiguity is detected in a request, it has been mainly addressed by asking the user clarification questions~\cite{Dogan2022,Pramanick2022,Chisari2025,Ren2023,Park2024,Abugurain2024,Liang2024}. For example, in~\cite{Thomason2015}, a semantic parser interpreted navigation and delivery task commands to ask clarification questions when the parser generated conflicting meanings. 
In both~\cite{Dogan2022} and~\cite{Pramanick2022}, image captioning networks compared user task descriptions with the robot's view to detect ambiguity in object identification~\cite{Dogan2022} and manipulation~\cite{Pramanick2022} tasks. The robot then formulated a question based on: 1) spatial relations between the candidate objects and known references 
~\cite{Dogan2022}, or 2) object attributes extracted from the image captions
~\cite{Pramanick2022}.

Recently, LLMs and VLMs have been used for resolving task ambiguities~\cite{Chisari2025,Ren2023,Park2024,Abugurain2024,Liang2024}. 
For example, request ambiguities have been detected by a fine-tuned VLM that evaluates the visual scene~\cite{Chisari2025}. Furthermore, ambiguous requests have been identified when an LLM generates inconsistent robot actions for the same request while the objects in the scene description are reordered~\cite{Park2024}. Clarification questions were then generated to obtain the missing details from a user~\cite{Chisari2025,Park2024}. 
In~\cite{Ren2023}, an LLM was prompted with the text descriptions of a robot's current observations, obtained using open-vocabulary object detection tools, to generate possible object rearrangement and manipulation tasks.
The robot then applied conformal prediction, using a previously evaluated task dataset and a user-specified success rate, to suggest safe options. This method was extended in~\cite{Liang2024} by prompting the LLM to also generate a reason for the safety and validity of options, using few-shot in-context learning from the robot's previous actions.
In~\cite{Abugurain2024}, a classifier was trained to determine ambiguity in vectors representing navigation instructions. Then, an LLM generated contextually relevant clarification questions and filled in preference-dependent information 
using conversations in the robot's memory~\cite{Abugurain2024}.

In a handful of papers, inference has also been used to decipher request ambiguities~\cite{Muthugala2016,Wan2025}. For example, in~\cite{Muthugala2016}, fuzzy logic was used to infer exact quantities for navigating through a domestic environment based on the spatial constraints of the current environment and the specific types of actions the robot can perform. In~\cite{Wan2025}, user intent was inferred for collaborative household cleanup tasks using transformer models based on text descriptions of the objects the user handled prior to a new instruction.

\vspace{-0pt}
\subsection{Robot Task Assistance Personalization}

Robot task personalization focuses on performing specific tasks to meet users’ needs, including how~\cite{Sadigh2017,Zhang2019,Wu2023,Wang2025,Han2025,Candon2026} or when~\cite{Patel2025} a task should be performed by a robot. Users often need to explicitly provide examples of their preferences through natural language~\cite{Wu2023,Patel2025,Han2025} or visual demonstrations~\cite{Zhang2019,Wang2025,Han2025}. Robots can also personalize their behaviors by observing users performing their own tasks~\cite{Candon2026}. 

In~\cite{Sadigh2017}, a reward function encoding a user's preferences for how a dynamic system should behave, e.g., a preferred driving style, was learned from the user's relative preferences between pairs of candidate driving trajectories. In~\cite{Zhang2019}, a Gaussian process latent variable model was used to personalize dressing assistance based on a user’s upper-body movement limitations and physical comfort zones. The user moved their restricted arm randomly during the data collection and the robot constructed a personalized user model~\cite{Zhang2019}. In~\cite{Candon2026}, a robot modeled user preferences, including preferred sub-task sequences and the number of items in a shared space, by passively observing the user performing their part of a shared task, such as making pizza, as implicit feedback. Maximum likelihood estimation was used to combine this feedback with explicit binary feedback obtained by pressing buttons on a table and to update user preference parameters~\cite{Candon2026}.

LLMs and VLMs have also been used to personalize robot task assistance. For example, in~\cite{Wu2023}, an LLM inferred general sorting rules from text examples of user preferences, for a robot to assist with household organizing tasks. 
In~\cite{Wang2025}, a VLM first generated text descriptions of visual user demonstrations of arranging objects. 
A Bayesian network was then used to ask clarification questions from the user to learn generalizable preferences~\cite{Wang2025}. 
In~\cite{Han2025}, an LLM task planner was aligned with a user's preferences for object placement by fine-tuning via imitation learning and iterative reinforced self-training. In~\cite{Patel2025}, an LLM represented household tasks by abstract concepts such as fragile objects; then a transformer model predicted how a new task should be handled, 
using a user's previous feedback on delegating tasks to a robot. 

\begin{figure*}[t]
    \centering
    \includegraphics[width=\linewidth]{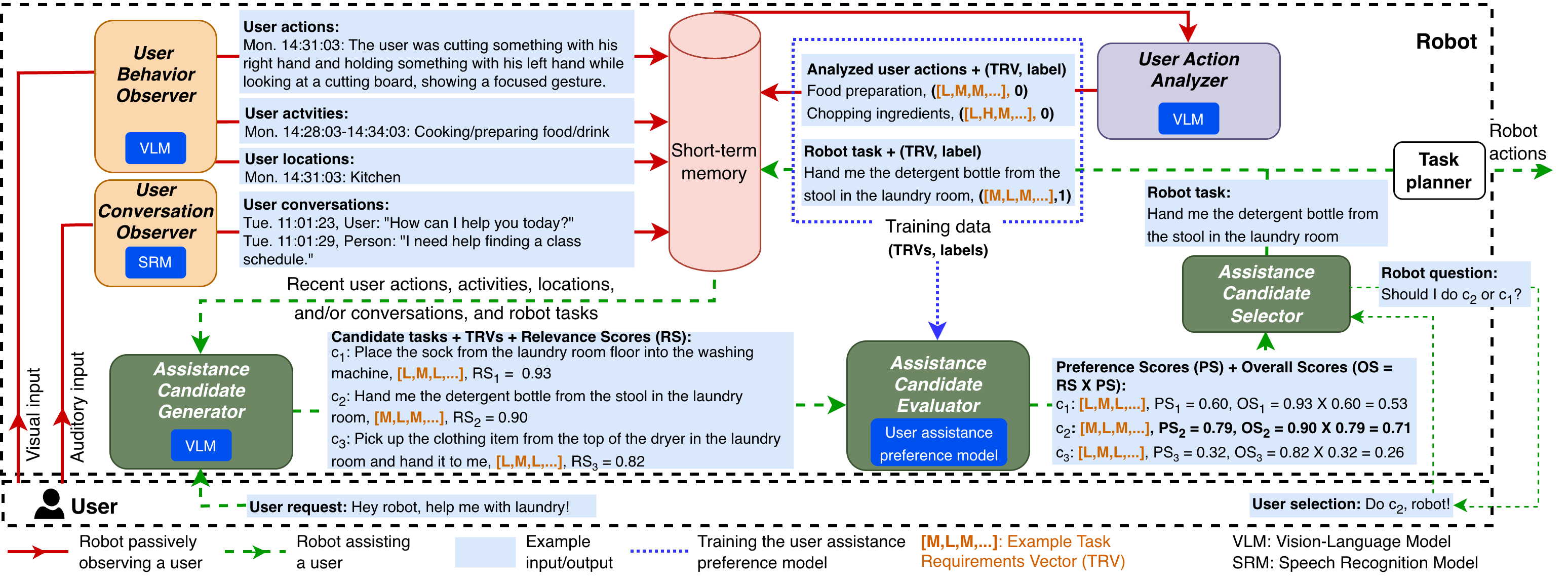}\vspace{-0pt}
    \caption{The PARAssist architecture. A robot passively observes a user via visual and auditory inputs and analyzes user actions (red lines). To assist based on a user request, the robot uses the stored user data and generates and evaluates multiple candidate tasks to assist (green dashed lines).}
    \label{fig:framework}
    \vspace{-0pt}
\end{figure*}

\subsection{Summary of Limitations}

Existing disambiguation methods primarily ground clarification questions and candidate tasks only in the present user request and  scene  ~\cite{Muthugala2016,Dogan2022,Pramanick2022,Chisari2025,Ren2023,Park2024,Liang2024}. 
A few methods have incorporated user history to: 1) avoid asking already-known details~\cite{Abugurain2024}, 2) disambiguate objects within a single interaction~\cite{Wan2025}, or 3) improve a robot's language parser across interactions~\cite{Thomason2015}. However, to the authors' knowledge, these methods do not use what a user has done recently to infer what assistance they would prefer. This can limit a robot's ability to ground disambiguation in user goals. 
Meanwhile, personalization methods mainly learn preferred assistance for defined tasks from explicit user demonstrations or feedback~\cite{Sadigh2017,Zhang2019,Wu2023,Wang2025,Han2025}. The exception is~\cite{Candon2026}
; however, with its focus on a single task, the learned user preferences cannot be easily transferred to other assistive tasks.
To address these limitations, PARAssist grounds request disambiguation in the history of user states and uses a user preference model for assistance across tasks. PARAssist reasons over real-world visual observations, rather than solely textual descriptions~\cite{Abugurain2024,Wan2025,Park2024}, which typically cannot capture how a specific user performs tasks in a particular environment (e.g., their reaching distance or lifting posture).

\vspace{-0pt}
\section{PARAssist Architecture}

The PARAssist architecture consists of six main modules, Fig.~\ref{fig:framework}.
Both the \textit{User Behavior Observer} and the \textit{User Conversation Observer} modules use vision and audio inputs to obtain user states. These user states are then provided to the \textit{User Action Analyzer} to generate Task Requirements Vectors~(TRVs) specifying physical and cognitive requirements of user actions (e.g., mobility or memory requirement). 

The \textit{Assistance Candidate Generator} utilizes the `user history,' including user conversations with others, previous actions related to the tasks, activities, and locations, as well as previous tasks the robot assisted with, to generate a set of candidate robot tasks, taking into account an ambiguous request. Each candidate task is paired with a relevance score specifying its relevance to the specific scenario context and user history, and a TRV estimating the task's requirements as if the user were to perform the task themselves. 
These TRVs are then evaluated by the \textit{Assistance Candidate Evaluator} to determine a preference score that would justify robotic assistance with each candidate task for a user. The \textit{Assistance Candidate Selector} then uses the two scores assigned to each candidate task to suggest the top candidates to the user. The user then selects the final task for the robot.
The description of the selected task is sent to a task planner to generate the robot actions. 
The task's TRV is stored in the short-term memory. The TRVs for both robot tasks and user actions are used as examples for training and updating the UAPM within the \textit{Assistance Candidate Evaluator} during deployment. The modules are detailed as follows:

\begin{figure}[t]
\centering
\includegraphics[width=\linewidth]{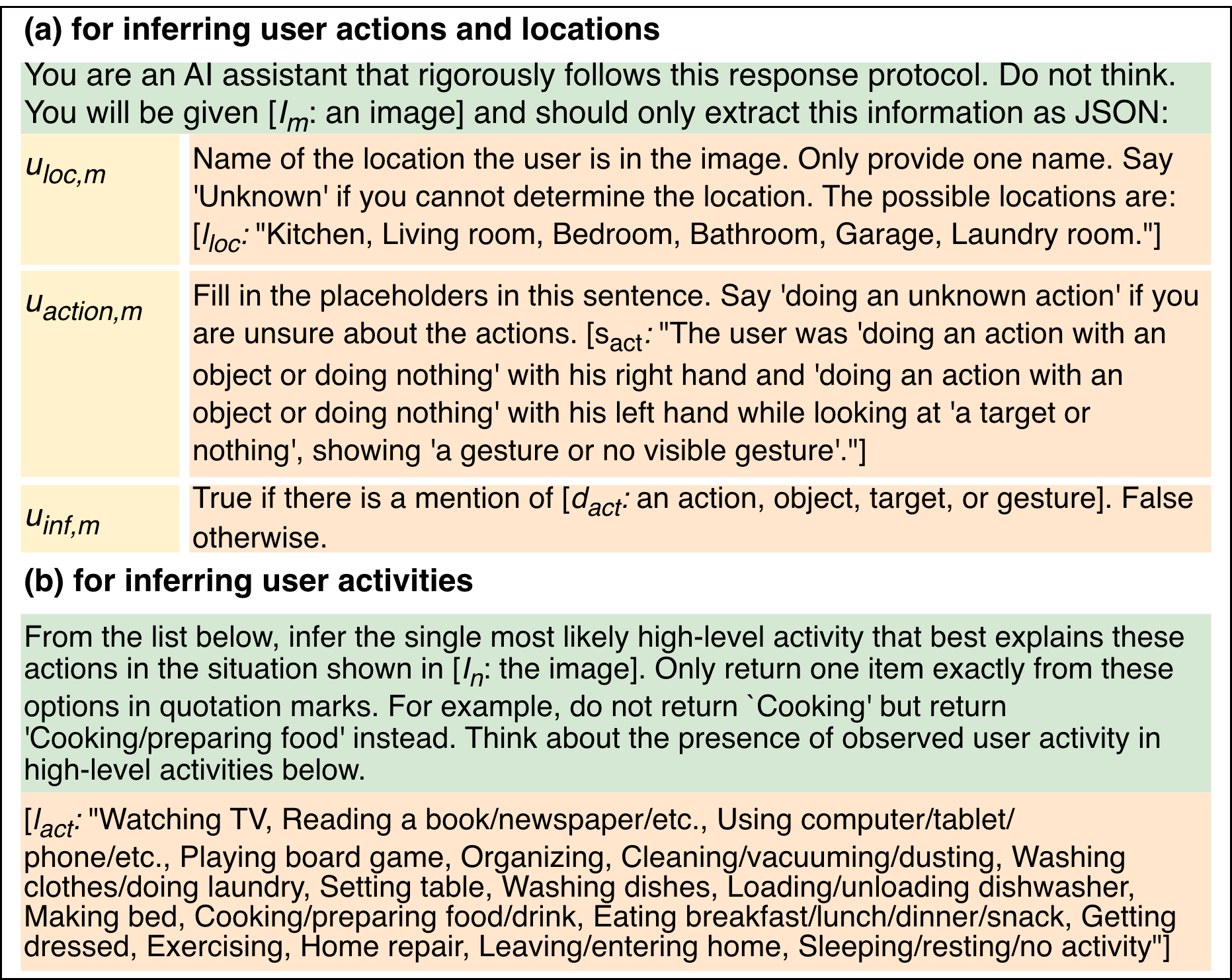}\vspace{-0pt}
\caption{VLM prompts in \textit{User Behavior Observer} for inferring: (a) user actions and locations, and (b) user activities. [\textit{Field names}: ``examples from our implementation in the domestic physical assistance domain''] are shown.}
\vspace{-15pt}
\label{fig:prompts1}
\end{figure}

\begin{figure}[h!]
\centering
\includegraphics[width=\linewidth]{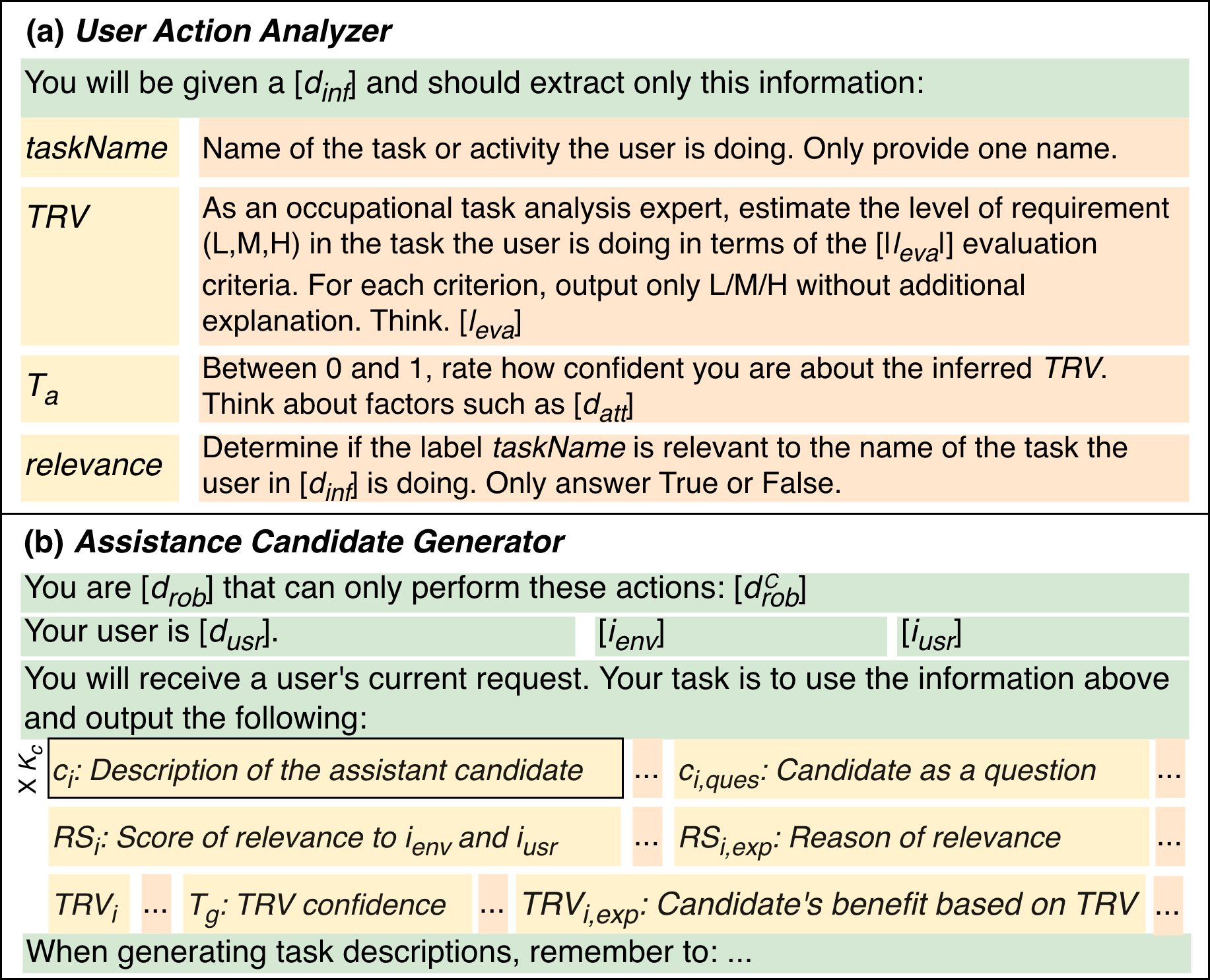}\vspace{-0pt}
\caption{VLM prompt structures in: (a) \textit{User Action Analyzer}, and (b) \textit{Assistance Candidate Generator}. \textit{Field names} are shown in brackets. Additional instructions for the VLM, which include $d_{usr}^{I}$, $d_{env}^{I}$, and $d_{tsk}$, are replaced with ... due to space constraints.}
\vspace{-0pt}
\label{fig:prompts2}
\end{figure}

The \noindent{\textit{User Behavior Observer}} module utilizes a lightweight VLM (VLM\#1) to infer user states, including actions, locations, and activities. The prompt to the VLM for inferring a user's actions and locations, Fig.~\ref{fig:prompts1}(a), includes an instruction to identify the user’s location $u_{loc,m}$ from a pre-defined list $l_{loc}$, describe any visible action $u_{action,m}$ by filling in placeholders in a pre-defined sentence $s_{act}$, and indicate by $u_{inf,m}$ whether $u_{action,m}$ contains any information ($m$ denotes observation date and time). The user actions and locations are periodically inferred, every $t_{obs}$ seconds, as behavior observation instances, using an input image frame $I_{m}$ from the robot's camera at $m$. 
$u_{action,m}$, $u_{inf,m}$, $I_{m}$, and $m$ are stored in the robot’s short-term memory. $u_{loc,m}$ is also added if $k_{loc}$ of the previous $n_{loc}$ inferences agree; otherwise, it is determined and recorded as an `unknown location.' This is to prevent generating assistance candidates based on noisy or incorrect inputs in the later stages.

The VLM prompt for inferring a user's activities, Fig.~\ref{fig:prompts1}(b), includes a text sequence of $u_{action,t}, \forall t: n-t_{hist}\leq t< n$ (i.e., user actions over the previous $t_{hist}$ seconds) and $I_{n}$ to infer user activity $u_{activity,n}$ (e.g., cooking and cleaning) from a list of available activities $l_{act}$ ($n$ denotes inference date and time). User activities are periodically inferred every $t_{act}$ seconds as activity inference instances. 
Presenting user actions as a series preserves their temporal relationships for activity inference. To reduce noise, the user activity output is smoothed using a majority voting sliding window of length $n_{act}$. If $k_{act}$ consecutive smoothed outputs indicate the same activity, the user activity is stored in the robot’s short-term memory with its start time, until this condition is no longer satisfied, at which point the end time is added.

The \noindent{\textit{User Conversation Observer}} module utilizes a Speech Recognition Model~(SRM) to transcribe conversations between a user and other people, and store them in the robot's short-term memory, with timestamps, to be used later for task disambiguation. Speech is associated with the user or other speakers using a speaker diarization method. 

The \noindent{\textit{{User Action Analyzer}}} 
module utilizes a VLM with reasoning token support (VLM\#2) to infer TRVs for user actions, based on either \textit{visual} or \textit{conversational} cues. The prompt for both modalities 
includes the following pre-defined fields for a given application domain such as domestic physical assistance, Fig.~\ref{fig:prompts2}(a): 1) description of what the VLM is prompted with for inferring TRVs ($d_{inf}$), 2) a list of criteria for evaluating user actions ($l_{eva}$), and 3) a description of important elements in $d_{inf}$ for this~inference~($d_{att}$).

When \textit{visual cues} are used, the \textit{User Action Analyzer} is activated after a period of the user performing an activity $a$ has been observed by the \textit{User Behavior Observer}. 
Let
$\mathcal{M}_{a} = \{\, m : t^{a}_{s} \le m \le t^{a}_{e},\ u_{inf,m}=\text{True} \,\}$
be the set of informative behavior observation instances during
activity $a$, where $t^{a}_{s}$ and $t^{a}_{e}$ are the start and end times. The set $\mathcal{S} \subseteq \mathcal{M}_{a}$, $|\mathcal{S}| = k_{a}$,
is chosen randomly. 
For each $m \in \mathcal{S}$, the VLM is prompted with the image frame $I_{m}$ to infer a TRV and a confidence score $T_{a} \in [0,1]$. The VLM also identifies the task the user appears to be performing (\textit{taskName}), and evaluates whether it is relevant to $a$ (\textit{relevance}). Only TRVs with $T_{a}$ larger than a given threshold $thr_{a}$ and relevant to the user activity are stored in the short-term memory. 

When \textit{conversational cues} are used, the \textit{User Action Analyzer} is activated after a new conversation between the user and other people is detected, bounded by two consecutive greeting keywords spoken by the user. The conversation transcript and task-relevant visual information (e.g., a map of the room) are provided to guide the TRV inference.

Once a user request $r$ is detected by the \textit{User Conversation Observer}, it is provided to the {\textit{Assistance Candidate Generator}} module to initiate the robot assistance process. This module compiles information about the user and the environment into a prompt, Fig.~\ref{fig:prompts2}(b), and queries VLM\#2
to generate a set of candidate tasks the robot can perform.
The 
 prompt 
has the following fields pre-defined for a given application domain: $A =\{i_{env}, i_{usr}\}$ is provided by the short-term memory, while $B=\{d_{rob}, d_{usr}, d_{rob}^{C}, d_{usr}^{I}, d_{env}^{I}, d_{tsk}\}$ is a set of constants. 
The elements are described as follows:

\begin{itemize}[leftmargin=*]
    \item $i_{usr}$, $i_{env}$: The \textit{user history} and the \textit{information about the environment} to inform generating assistance candidates. 
    \item $d_{usr}$: A short description about the {user}. 
    \item $d_{rob}$, $d_{rob}^{C}$: The \textit{robot's short description} and \textit{its capabilities} for assistance in the application domain.
    \item $d_{usr}^{I}$, $d_{env}^{I}$: Description of 
    $i_{usr}$ and $i_{env}$.
    \item $d_{tsk}$: Elements required to clearly define tasks.
\end{itemize}

The VLM generates the descriptions of $k_{c}$ assistance candidates $c_{i} \in C$ given $A \cup B \cup \{r\}$ and also provides each candidate in the form of a question $c_{i,ques}$ to the user for communication purposes, e.g., ``\textit{Do you want me to bring you a cup?}" For each candidate $c_{i}$, the VLM also outputs a relevance score to $i_{usr}$ and $i_{env}$ ($RS_{i}$) and an explanation for why it is suggested ($RS_{i,exp}$). This explanation is stated from the robot’s perspective, and may include evidence from $d_{usr}^{I}$, $d_{env}^{I}$, and times (e.g., ``\textit{Because I saw you in the laundry room two minutes ago.}"). 
The VLM also generates a TRV for each candidate ($TRV_{i}$), based on the same criteria used by the \textit{User Action Analyzer} (i.e., $l_{eva}$), as if the user were to perform the task, along with a confidence score $T_{g} \in [0,1]$. If any aspects of the task demands are rated high, the VLM also generates an explanation $TRV_{i,exp}$ of how the candidate task can benefit the user, citing those particular aspects, e.g., ``\textit{It can save you from lifting heavy objects.}" If the user request clearly defines a task with $d_{tsk}$, the VLM outputs will only be related to the user request (i.e., $|C|=1$ and $c_{1}=r$).

The generated assistance candidates are provided to the {\textit{Assistance Candidate Evaluator}} module, which includes the UAPM, a probabilistic classifier that estimates a preference score that the user would prefer robotic assistance with a candidate task, given its TRV. The UAPM inputs the TRV for each generated assistance candidate $c_{i}$ and assigns a preference score $PS_{i} \in [0,1]$. Overall scores $OS_i = RS_i \times PS_i$ are determined by combining the relevance scores and preference scores of the assistance candidates. 
Once the overall scores are available, the {\textit{Assistance Candidate Selector}} module suggests the $k_{top}$ candidate(s) with the highest overall score(s) 
to the user by stating $c_{i,ques}$, optionally with $RS_{i,exp}$ and/or $TRV_{i,exp}$. The text description of the candidate chosen by the user (i.e., $c_{i}$) is then sent to the task planner to generate appropriate robot actions.

\section{Training}

The UAPM is trained and updated on the TRVs from tasks the user performed independently (label 0, inferred by the \textit{User Action Analyzer}) and those for which the user requested robotic assistance (label 1, generated by the \textit{Assistance Candidate Generator}), before each new request. 
To reduce bias toward class 0 when the user performs many tasks themselves and requests robot assistance for only a few, random undersampling of label 0 TRVs is utilized~\cite {Aguiar2024}, while retaining all label 1 TRVs. This ensures all analyzed user behavior observation instances can be included in training. Training is performed using TRVs from the past $k_{d}$ days in the robot’s short-term memory for adaptation to user preferences over time, e.g., with changes in physical condition.

\vspace{-0pt}
\section{Assistance Domains}

We implemented the PARAssist architecture for assistance with: 1) domestic tasks in a home, and 2) addressing inquiries to a receptionist in a university. We chose these domains to examine personalization in disambiguating based on both visual and conversational cues. We used OpenAI GPT-5.4~\cite{openai2026} as VLM\#2. We also used an NVIDIA GeForce RTX 3090 24 GB GPU to host VLM\#1 and the SRM, and an Intel Core Ultra 7-265 CPU to train the UAPM.

\vspace{-0pt}
\subsection{Assistance Domain 1: Domestic Assistance}
\label{phys_imp}

The domestic assistance domain involves a service robot providing support to a user with everyday physical tasks at home, such as retrieving objects. 
The \textit{User Behavior Observer} uses real-time RGB camera input or a series of offline videos, each with a start date and time, to observe the user. YOLO11n~\cite{Jocher2024} is used to detect the user in these inputs. The Qwen3.5-9B VLM~\cite{Qwen35} is used as VLM\#1, with $t_{obs}=2.5$
, $t_{act}=25$, $t_{hist}=60$, $k_{loc}=k_{act}=3$, and $n_{loc}=n_{act}=5$. 
The \textit{User Action Analyzer} uses the \textit{visual} modality with $k_{a}=5$ and $thr_{a}=0.7$.
$d_{inf}$ and $d_{att}$ are `an image of the user' and `visibility of objects and the user, properties and locations of objects, etc.' $l_{eva}$ consists of 15 criteria in standardized functional and cognitive capacity evaluation scales, which assess physical and cognitive ability for work and living~\cite{WorkSafeBC2023,WashingtonLNI2016}. The VLM rates each criterion as H~(high), M~(medium), or L~(low). The criteria consist of: Gross motor strength~\footnote{We also noted `give L for picking less than around 1~kg, M for 1--2~kg, and H for heavier loads,' for improved VLM inference consistency.}, reaching overhead, bending down, fine motor control, grip strength, near vision, far vision, depth perception, color and texture discrimination, planning/organizing, memory, attention, static posture endurance, temperature tolerance, and slip and trip risk tolerance.

For the \textit{Assistance Candidate Generator}, $i_{env}$ consists of: 1) a list of rooms in the house as noted in Fig.~\ref{fig:prompts1}(a), 2) an explanation that `the objects in the current location are those visible in the image,'
 and 3) the robot's current location (same as the user's room).
$i_{usr}$ includes user actions in the past 5 minutes, user activities in the past 7 days, and the user's last 5 requests from the robot, with the current date and time. These windows are selected to capture immediate context, weekly habitual patterns, and recent assistance preferences. The pre-defined prompt fields are ($k_{c}=10$):

\begin{itemize}[leftmargin=*]
    \item $d_{usr}$: `a resident in the house'
    \item $d_{rob}$: `a mobile service robot with arms in a house'
    \item $d_{rob}^{C}$ defines the robot’s available low-level capabilities, including localization, room and object detection, navigation, door operation, object pickup, placement, and handover.
    \item $d_{usr}^{I}$: `relevant and recent user activities, relevant user tasks in the past week that may indicate a pattern, and relevant and recent user requests, all with their time of the day'
    \item $d_{env}^{I}$: `possible locations, objects in this location, and the environment state observed in the image'    
    \item $d_{tsk}$: `objects and locations'
\end{itemize}

The UAPM is a logistic regression model, chosen for its simplicity in model formulation and suitability for small training sets, to reduce the risk of overfitting~\cite{Cawley2010}. The maximum number of iterations is set to 1000, and $k_{d}=7$. 
The \textit{Assistance Candidate Selector} sends the assistance task candidates to the few-shot task planner CodeBotler~\cite{Hu2024}, which uses OpenAI GPT-5.4 with `medium' reasoning effort and the robot's low-level functions based on $d_{rob}^{C}$.

\subsection{Assistance Domain 2: University Receptionist Assistance}
\label{pepper_imp}

\begin{figure*}[htp!]
\vspace{-0pt}
    \centering
    \includegraphics[page=1,width=\linewidth]{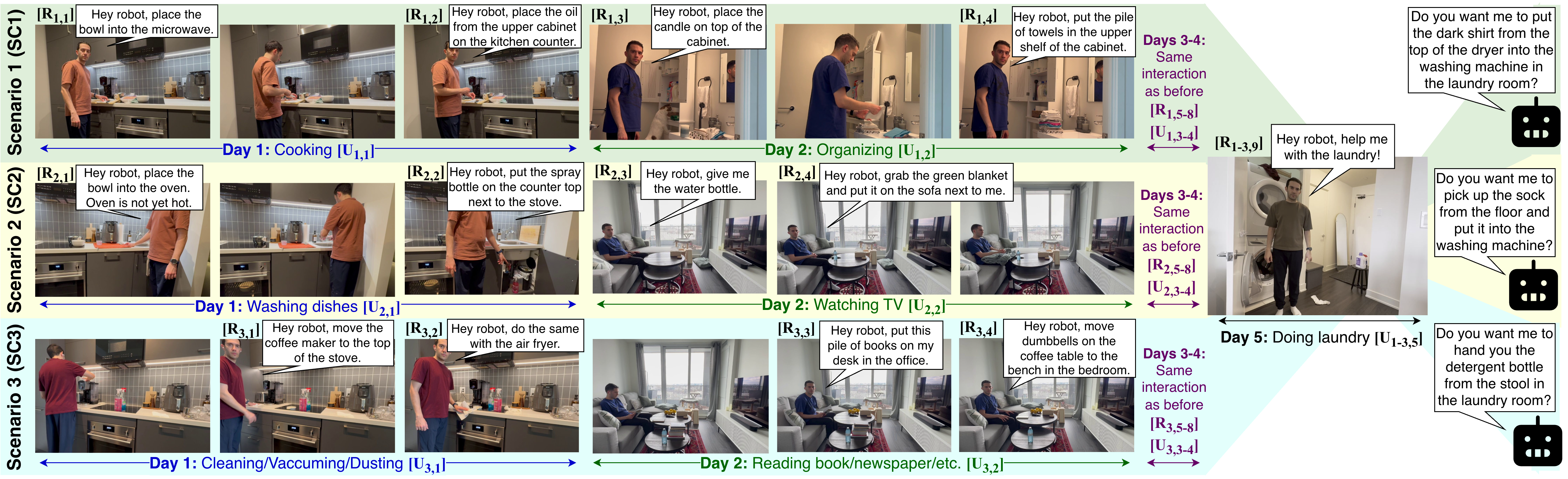}\vspace{-0pt}
    \caption{Interaction of a user with the PARAssist architecture. $[R_{i,j}]$ and $[U_{i,j}]$ denote the $j$-th user {request} and {activity} in the $i$-th scenario. The most frequently suggested assistance candidate in the top two suggestions across 10 trials in the MRE condition is shown on the far right.}
    \label{fig:domestic}
    \vspace{-0pt}
\end{figure*}

The university receptionist assistance domain involves a social robot 
supporting a university receptionist with student and instructor assistance tasks. 
The \textit{User Conversation Observer} monitors conversations between the receptionist and students or instructors. It performs real-time speech transcription by segmenting microphone audio streams into utterances using energy thresholding and voice activity detection~\cite{Bain2022} (each segment is separated by 1 s of no speech input). Each segment is transcribed with WhisperX {large-v3}~\cite{Bain2022}. Speaker diarization is performed using WhisperX’s \textit{DiarizationPipeline}, which compares each voice embedding with an existing embedding of the receptionist’s voice. 
The \textit{User Action Analyzer} uses the \textit{conversational} modality, with a `high' VLM reasoning effort. 
Conversations are segmented by greeting keywords such as ``\textit{hi}.'' $d_{inf}$ and $d_{att}$ are `a conversation and the building floor layout' and `locations, school policy titles, school items, lectures, or times.' $l_{eva}$ consists of: Walking, school items inventory memory, school class schedule memory, and school policy knowledge; to rate as H~(high) or L~(low). A building floor layout with dimensions is included in the prompt, with a walking criterion of: `give H for more than 10 m and L for less than 10~m.' 

The \textit{Assistance Candidate Generator} is triggered by the receptionist saying ``\textit{hey robot}'' and a request, with a `medium' VLM reasoning effort. $i_{env}$ states that the robot is at the reception and `the possible locations are those observed in the layout of the floor in the image.' $i_{usr}$ is the most recent piece of user conversation. The \textit{Assistance Candidate Evaluator} and the UAPM are similar to those used for the domestic assistance domain. The prompt fields are ($k_{c}=3$):

\begin{itemize}[leftmargin=*]
    \item $d_{usr}$: `a university receptionist'
    \item $d_{rob}$: `a mobile social robot in a university'
    \item $d_{rob}^{C}$ includes tracking time and location, retrieving lecture and classroom schedules, navigation to specified locations, checking room occupancy, and verbal announcements.
    \item $d_{usr}^{I}$: `recent user conversations with students/instructors'
    \item $d_{env}^{I}$: `possible locations'    
    \item $d_{tsk}$: `locations, lectures, and times'
\end{itemize}

The \textit{Assistance Candidate Selector} allows the robot to provide $c_{i,ques}$ for the top candidate task ($k_{top}=1$). If the user agrees, the robot task is sent to the planner~\cite{Hu2024}.
The robot's RGB camera is used to determine 
room occupancy, by sending an image to the GPT-5-nano VLM with the prompt 
‘\textit{Is anyone visible in the room through the doorway?}’

\section{Domestic Assistance Scenarios}

We conducted: 1) trials to evaluate PARAssist handling ambiguous requests in multiple scenarios and with various VLM reasoning efforts, and 2) an ablation study to assess the role of user history and the UAPM in disambiguation. 

In three scenarios (SC1, SC2, and SC3), PARAssist handled user requests for performing physical tasks at home, Fig.~\ref{fig:domestic} ($i=1,2,3$). 
For each scenario, PARAssist received five videos of a user performing real-world activities. The first four videos ($[U_{i,1-4}]$) varied across the scenarios and consisted of two distinct activities, each repeated twice to extend the user observation period for training the UAPM. 
The user made eight unambiguous requests ($[R_{i,1-8}]$) during these activities. 
The fifth video ($[U_{1-3,5}]$) in all scenarios consisted of the user entering the laundry area, opening the washer door, pressing buttons on the washer, and making the ambiguous request ``\textit{help me with the laundry}" ($[R_{1-3,9}]$). 

\subsection{Personalized Assistance Response to Ambiguous Requests}
\label{exp2}

For each scenario, we conducted 10 trials with the full PARAssist framework in two conditions: 1)~Medium/high Reasoning Effort~(MRE): the VLM reasoning effort was `medium' in the \textit{Assistance Candidate Generator} and `high' in the \textit{User Action Analyzer}, and 2)~Low/medium Reasoning Effort~(LRE) was set for these two modules, respectively. 

The \textit{User Behavior Observer} was able to identify all user activities, except for organizing cleaning supplies ($[U_{1,2}],[U_{1,4}]$) in one trial, where it was incorrectly classified as a cleaning activity. 
Table~\ref{tab_trv} presents the  
TRVs in the MRE condition. In SC1, the user performed actions that did not require reaching high, while requesting assistance with tasks that required reaching high. 
In SC2, the user did not bend down in their actions; however, they requested assistance when bending was needed. 
In SC3, the user handled light lifting and requested assistance with heavy lifting.

The top two suggested tasks in response to the final user request in each trial (i.e., $s_1$ and $s_2$) were analyzed. Suggestions that were infeasible or unreasonable for the laundry task were flagged by the research team. For example, moving the detergent bottle to the top of the washer was infeasible as it was a stacked dual system, and taking clothes out of the dryer was unreasonable when it was empty. The remaining suggestions were evaluated based on two criteria, whether they: 1) included the \textit{key} aspect of reaching high for SC1, bending down for SC2, or heavy lifting for SC3; and 2) contributed to the \textit{user task} of starting the washer. 

\begin{figure}[tp!]
    \centering
    \includegraphics[width=\linewidth]{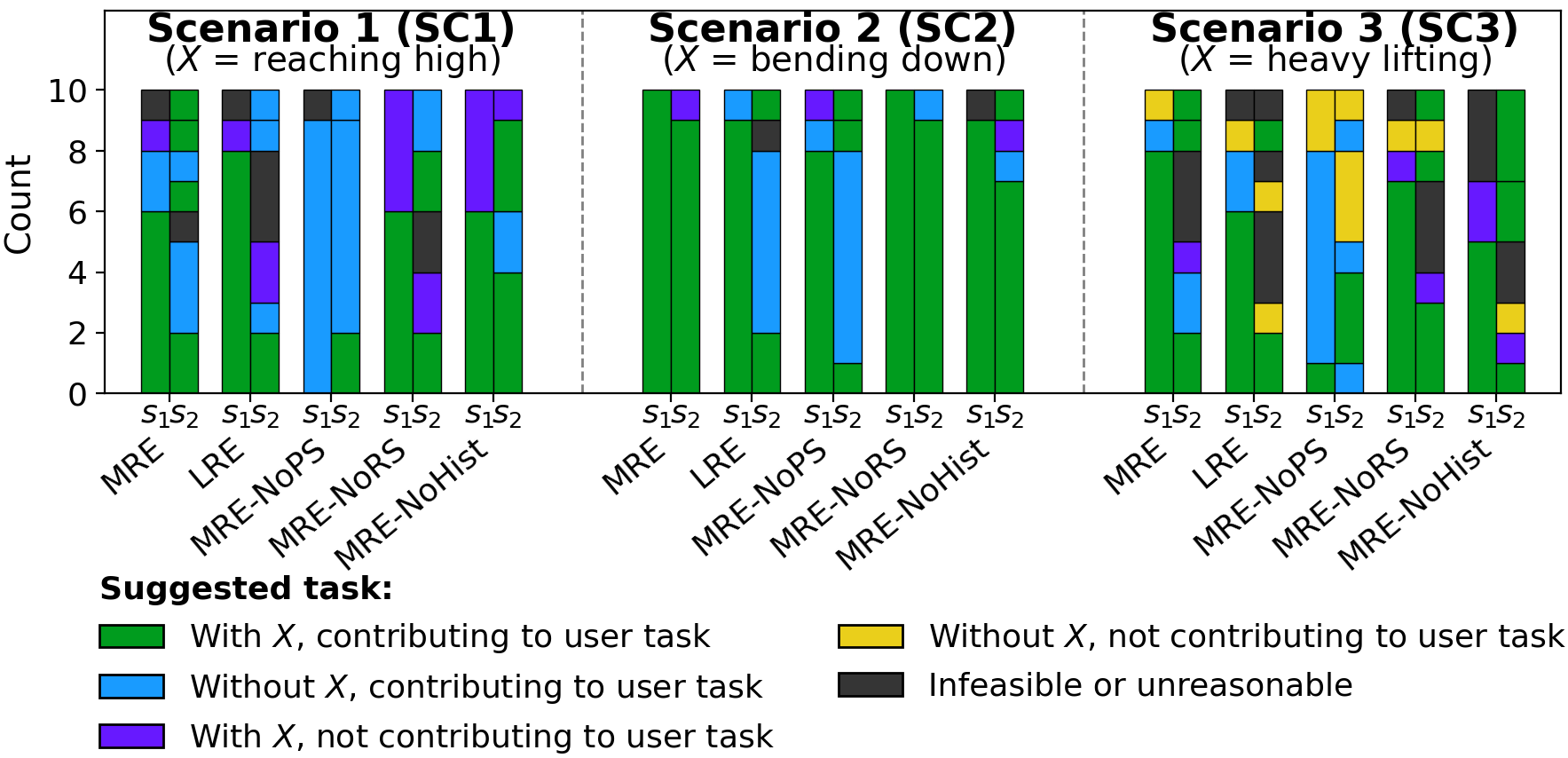}\vspace{-0pt}
    \caption{The top two suggested assistance candidates $s_1$ and $s_2$ for 10 trials, across multiple scenarios and conditions. X denotes the \textit{key} aspect.}
    \label{fig:domestic_chart}
    \vspace{-0pt}
\end{figure}

In both MRE and LRE conditions, the most frequently suggested candidate among $s_1$ and $s_2$ was to assist with: moving the clothing item from the top of the dryer (SC1), picking up the sock from the floor (SC2), and retrieving the laundry detergent from the stool (SC3). As shown in Fig.~\ref{fig:domestic_chart}, at least one of the top two candidate tasks met both criteria in 9, 10, and 10 trials under the MRE condition, and in 8, 10, and 7 trials under the LRE condition for SC1, SC2, and SC3, respectively. Reducing the reasoning effort of the VLMs led to a lower performance, potentially due to reduced accuracy in candidate generation (more infeasible suggestions for SC1 and SC3) or in TRV estimation (more suggestions lacking the \textit{key} aspect for SC2). 
The results also varied across scenarios. In SC2, suggestions often included bending down, likely due to the sock on the floor being a visually prominent object. 
In SC3, suggestions were sometimes infeasible, possibly due to the VLM not being able to accurately estimate object weights from visual cues.

\begin{table}[tp!]
\caption{Mean (and SD) of TRVs (first three elements) for user $[U_{i,j}]$ and robot $[R_{i,j}]$ tasks in the MRE condition.}
\vspace{-0pt}
\label{tab_trv}
\centering
\ssmall
\setlength{\tabcolsep}{2pt}
\vspace{-0pt}
\begin{threeparttable}
\begin{tabular}{@{}p{0.4cm}p{2.6cm}p{2.6cm}p{2.6cm}@{}}
\toprule
 & \multicolumn{1}{c}{\textbf{SC1} ($i=1$)} & \multicolumn{1}{c}{\textbf{SC2} ($i=2$)} & \multicolumn{1}{c}{\textbf{SC3} ($i=3$)} \\ \midrule
$U_{i,1}$ & {[}0.0(0.0), \textbf{0.0(0.1)}, 0.1(0.2), ...{]} & {[}0.0(0.0), 0.0(0.0), \textbf{0.0(0.1)}, ...{]} & {[}\textbf{0.0(0.0)}, 0.0(0.0), 0.0(0.0), ...{]} \\
$U_{i,2}$ & {[}0.0(0.1), \textbf{0.3(0.5)}, 0.0(0.2), ...{]} & {[}0.0(0.0), 0.0(0.0), \textbf{0.0(0.0)}, ...{]} & {[}\textbf{0.0(0.0)}, 0.0(0.0), 0.0(0.0), ...{]} \\ \cmidrule(l){2-4}
$R_{i,1}$ & {[}0.2(0.4), \textbf{2.0(0.0)}, 0.0(0.0), ...{]} & {[}0.1(0.2), 0.0(0.0), \textbf{1.9(0.4)}, ...{]} & {[}\textbf{1.6(0.5)}, 0.0(0.0), 0.0(0.0), ...{]} \\
$R_{i,2}$ & {[}0.0(0.0), \textbf{2.0(0.0)}, 0.0(0.0), ...{]} & {[}0.0(0.0), 0.0(0.0), \textbf{1.7(0.6)}, ...{]} & {[}\textbf{1.9(0.4)}, 0.0(0.0), 0.0(0.0), ...{]} \\
$R_{i,3}$ & {[}0.0(0.0), \textbf{2.0(0.0)}, 0.0(0.0), ...{]} & {[}0.0(0.0), 0.0(0.0), \textbf{1.9(0.5)}, ...{]} & {[}\textbf{1.1(0.4)}, 0.0(0.0), 0.4(0.5), ...{]} \\
$R_{i,4}$ & {[}1.0(0.0), \textbf{2.0(0.0)}, 0.0(0.0), ...{]} & {[}0.0(0.0), 0.0(0.0), \textbf{1.9(0.4)}, ...{]} & {[}\textbf{1.8(0.4)}, 0.0(0.0), 0.5(0.5), ...{]} \\ \bottomrule
\end{tabular}
\begin{tablenotes}
\footnotesize
\item H, M, and L are converted to numerical values 2, 1, and 0, respectively.
\end{tablenotes}
\end{threeparttable}
\vspace{-0pt}
\end{table}

The average latency of the \textit{Assistance Candidate Generator} for the MRE condition was 75.5~s (SD = 17.3), 56.2~s (SD = 12.3), and 62.3~s (SD = 8.8) for SC1, SC2, and SC3, respectively. For the LRE condition, latency was 23.6~s (SD = 6.7), 22.4~s (SD = 1.0), and 20.4~s (SD = 2.8), respectively, meaning that reducing the reasoning effort led to a reduction in system latency.

\vspace{-0pt}
\subsection{Ablation Study}

We conducted an ablation study to explore the contribution of individual design choices in the \textit{Assistance Candidate Generator} and \textit{Assistance Candidate Evaluator}. We compared MRE with three of its variations, Fig.~\ref{fig:domestic_chart}: 1) MRE-NoPS, where $PS=1$ for all candidates (i.e., the \textit{Assistance Candidate Evaluator} was removed); 2) MRE-NoRS, where $RS=1$ for all candidates and thus, candidates were not ranked by relevance to user history; and 3) MRE-NoHist, where $d_{usr}^{I} = \varnothing$ and user history was excluded in candidate generation. 
For the MRE-NoPS condition, none of the top two candidate tasks included the \textit{key} aspect in 7 and 6 
trials for SC1 and SC3. The reduced number of task suggestions with the \textit{key} aspect emphasizes the need for user-specific preference scoring for personalized disambiguation, as candidate relevance to the user history alone may not capture individual differences in assistance needs. For both MRE-NoRS and MRE-NoHist variants for SC1, the first suggestions in 4 trials involved tasks that did not contribute to starting the washer, even though these suggestions included the \textit{key} aspect. For the MRE-NoHist variant for SC3, the first suggestions in 3 trials were infeasible or unreasonable, and in 2 trials did not contribute to starting the washer. Without considering user history (MRE-NoRS and MRE-NoHist variants), the system suggested tasks that did not contribute to the \textit{user task}, and could therefore be unhelpful. Increased infeasible or irrelevant suggestions for SC3 may have occurred as the dryer was also visible to the system, which could have contributed to the VLM not being able to infer the user's intent from the scene alone.

\vspace{-0pt}
\section{Real-World Experiments}
\label{exp3}

\begin{figure}[tp!]
    \centering
    \includegraphics[width=\linewidth]{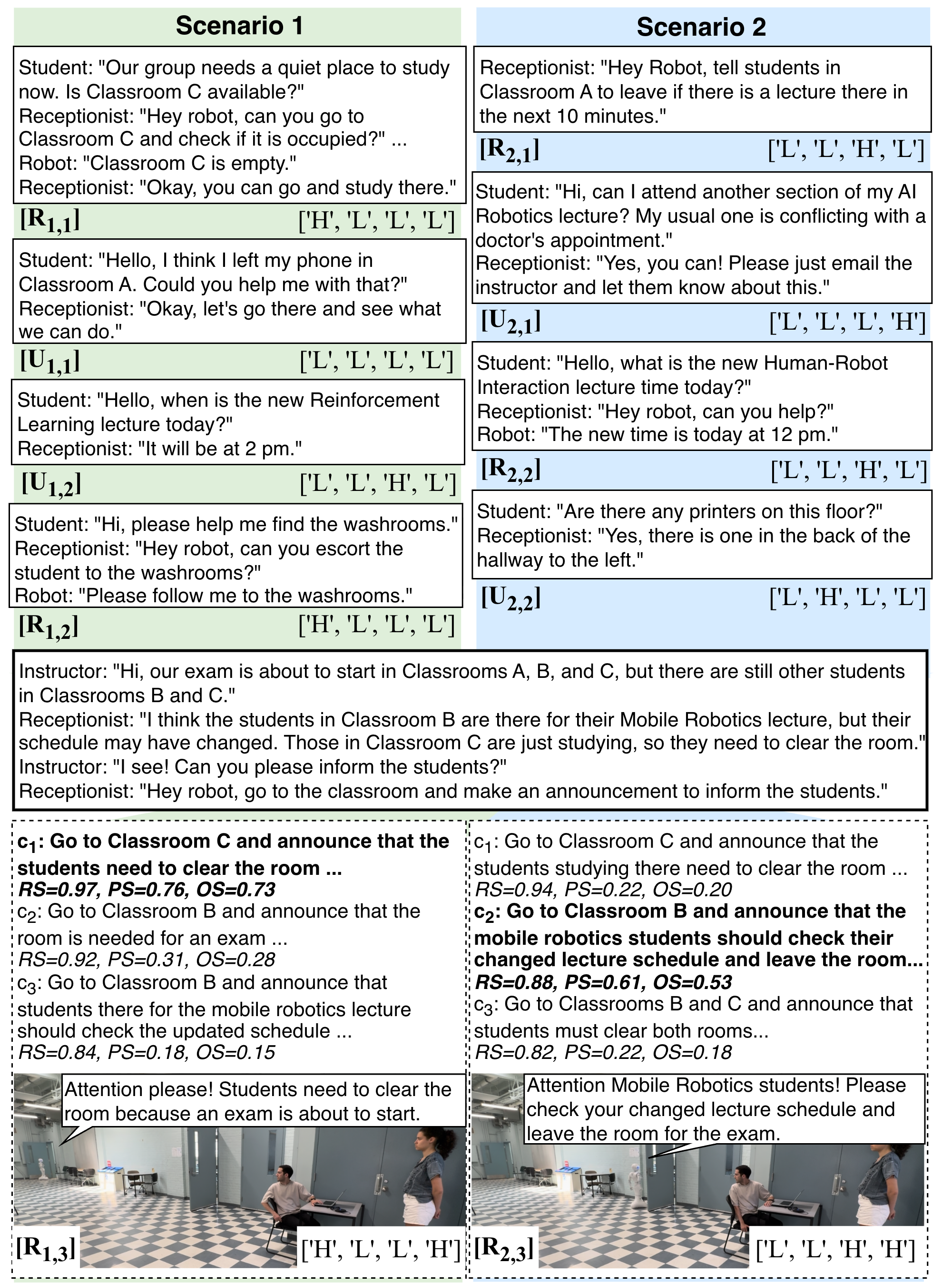}\vspace{-0pt}
    \caption{Interaction of a receptionist with students, an instructor, and a robot. Each box indicates a conversation labeled in the bottom left. $R_{i,j}$ denotes the $j$-th user request from the robot and $U_{i,j}$ denotes the $j$-th user task in scenario $i$. The TRV  for each user action or robot task is shown in the bottom-right corner. The robot assisted with the task represented in bold after the last user request. Relevance ($RS$), preference ($PS$), and overall scores ($OS$) for each candidate are included.}
    \label{fig:university}
    \vspace{-0pt}
\end{figure}

We conducted a set of experiments in a real-world environment to investigate the performance of the PARAssist embedded in a physical robot. In these experiments, a member of the research team was a university receptionist, while other research group members were the students and an instructor. In two scenarios ($i=1,2$), the receptionist directly assisted students twice ($[U_{i,1-2}]$) and requested the robot's assistance for supporting students twice ($[R_{i,1-2}]$), Fig.~\ref{fig:university}. The TRV for each user action and robot task is presented in Fig.~\ref{fig:university}. In Scenarios 1 and 2, the receptionist asked for robotic assistance only when the task required long-distance walking or class schedule memory, respectively. 

Both scenarios concluded with a conversation about two classrooms needed for an exam being unexpectedly occupied for different reasons ($[R_{i,3}]$). In response to the receptionist’s request for the robot to make an announcement, which was ambiguous as the classroom was not specified, the robot provided assistance consistent with recent patterns in the receptionist's requests and actions. In Scenario 1, the robot made a general announcement in Classroom~C, which was far from the reception and required a long walk. 
In Scenario~2, the robot announced a schedule update, which required class schedule memory, in Classroom~B. These experiments demonstrated that PARAssist can reason from conversational cues and suggest personalized assistance aligned with a user's assistance preferences. A video demonstrating the capabilities of PARAssist for the receptionist assistance application is presented at our project webpage: \href{https://parassist.github.io}{https://parassist.github.io}, and on the ASBLab YouTube channel: \href{https://youtu.be/2IUEYxDudvE}{https://youtu.be/2IUEYxDudvE}.

\vspace{-0pt}
\section{Discussion}

PARAssist can enable service robots to handle ambiguous requests in a personalized way by combining context-aware reasoning with a learned user assistance preference model. PARAssist was able to reason over a history of user actions (shown in domestic assistance scenarios) or conversations (shown in receptionist assistance experiments) to infer assistance candidates, rather than relying primarily on the current request and scene, such as in~\cite{Pramanick2022,Park2024,Ren2023,Liang2024,Chisari2025}. PARAssist was also able to determine suitable assistance by learning user assistance preferences from the inferred TRVs of user actions and robot tasks, rather than from explicit user-provided examples, such as in~\cite{Wu2023,Wang2025,Han2025}. 

Limitations of PARAssist include periodic VLM inference in the \textit{User Behavior Observer}, which may miss fast user actions such as pressing buttons on the washer. As a result, PARAssist may suggest a task to assist with that the user has already completed.
To address this, the user observation pipeline could allocate more computational resources for faster inference or buffer captured image frames for future processing. Infeasible or unreasonable assistance generated by the \textit{Assistance Candidate Generator}, which was shown in the domestic assistance scenarios for the laundry task, could be mitigated by incorporating richer contextual representations such as semantic maps or scene graphs~\cite{Mohammadi2025}, and more detailed models of robot capabilities such as which objects can be manipulated. 

\section{Conclusion}

We developed the PARAssist framework for personalized and adaptive robotic assistance from ambiguous user requests. Our novel framework uniquely integrates passive and multimodal user observations and a user assistance preference model into a disambiguation pipeline to determine robot assistance. 
Future work will include extending the architecture to be generalized to new assistive application domains and conducting user studies to assess user acceptance and the appropriateness of disambiguation relative to a user's actual intent.

\section*{ACKNOWLEDGMENT}

The authors would like to thank the following members of the ASBLab at the University of Toronto: Sujith Santharuban and Abi Nevo for their assistance with conducting the experiments; and Haitong Wang, Matthew Lisondra, and Glenn Takashi Shimoda for their insightful discussions.

\bibliography{bibliography}
\bibliographystyle{IEEEtran}

\end{document}